\documentclass[conference]{IEEEtran}
\IEEEoverridecommandlockouts

\usepackage{algorithmicx}
\usepackage{makecell}
\usepackage{algorithm}
\usepackage{algpseudocode}
\algtext*{EndWhile}
\algtext*{EndIf}
\algtext*{EndFor}
\algtext*{EndProcedure}
\usepackage{amsmath}
\usepackage{amssymb}

\usepackage{array}
\usepackage{booktabs}
\usepackage[bookmarks=false]{hyperref}
\usepackage[noadjust]{cite}
\usepackage{caption}
\usepackage{color}
\usepackage{comment}
\usepackage[inline]{enumitem}
\usepackage{epsfig}
\usepackage{etoolbox}
\usepackage{fancybox}
\usepackage{float}
\usepackage{fixltx2e}
\usepackage{graphicx}
\usepackage{hyperref}
\usepackage{listings}
\usepackage{multirow}
\usepackage{multicol}
\usepackage{placeins}
\usepackage{rotating}
\usepackage{setspace}
\usepackage[hang, tight]{subfigure}
\usepackage{threeparttable}
\usepackage{times}
\usepackage{url}
\usepackage{verbatim}
\usepackage{wrapfig}
\usepackage[usenames,dvipsnames]
{xcolor}

\newbool{inccomment}
\setbool{inccomment}{true}
\newcommand{\YC}[1]{\ifbool{inccomment}{{\color{magenta}YC\@: #1}}{}}
\newcommand{\XC}[1]{\ifbool{inccomment}{{\color{blue}XC\@: #1}}{}}
\newcommand{\TD}[1]{\ifbool{inccomment}{{\color{orange}#1}}{}}
\newcommand{\FN}[1]{\ifbool{inccomment}{{\color{OliveGreen}#1}}{}}
\newcommand{\GR}[1]{\ifbool{inccomment}{{\color{Tan}#1}}{}}
\newcommand{\LD}{\ifbool{inccomment}{{\color{magenta}\\============================================\\}}}
\newcommand{\RF}{\ifbool{inccomment}{{\color{green}~[R]}}}

\newcommand{\roma}[1]{\uppercase\expandafter{\romannumeral #1\relax}}

\def\BibTeX{{\rm B\kern-.05em{\sc i\kern-.025em b}\kern-.08em
    T\kern-.1667em\lower.7ex\hbox{E}\kern-.125emX}}
    
\graphicspath{{}}

\begin{document}

\title{\vspace{-3mm}Helios: Heterogeneity-Aware Federated Learning with Dynamically Balanced Collaboration
\vspace{-5mm}
}


\author{\IEEEauthorblockN{Zirui Xu\IEEEauthorrefmark{1},
Fuxun Yu\IEEEauthorrefmark{1}, Jinjun Xiong\IEEEauthorrefmark{2} and
Xiang Chen\IEEEauthorrefmark{1}}
\IEEEauthorblockA{\IEEEauthorrefmark{1}George Mason University, \IEEEauthorrefmark{2}IBM Thomas J. Watson Research Center\\
Email: \IEEEauthorrefmark{1}$\{$zxu21, fyu2, xchen26$\}$@gmu.edu, 
\IEEEauthorrefmark{2}jinjun@us.ibm.com
\vspace{-5mm}
}}


\maketitle

\begin{abstract}

	As Federated Learning (FL) has been widely used for collaborative training, a considerable computational \textit{straggler} issue emerged: when FL deploys identical neural network models to heterogeneous devices, the ones with weak computational capacities, referred to as stragglers, may significantly delay the synchronous parameter aggregation.
	Although discarding stragglers from the collaboration can relieve this issue to a certain extent, stragglers may keep unique and critical information learned from non-identical dataset, and directly discarding will harm the overall collaboration performance.
Therefore, in this paper, we propose \textit{Helios} --- a heterogeneity-aware FL framework to tackle the straggler issue.
	\textit{Helios} identifies individual devices' heterogeneous training capability, and therefore the expected neural network model training volumes regarding the collaborative training pace.
	For straggling devices, a ``soft-training'' method is proposed to dynamically compress the original identical training model into the expected volume through a rotating neuron training approach.
	With extensive algorithm analysis and optimization schemes, the stragglers can be accelerated while retaining the convergence for local training as well as federated collaboration.
	Experiments show that \textit{Helios} can provide up to 2.5$\times$ training acceleration and maximum 4.64\% convergence accuracy improvement in various collaboration settings.

\end{abstract}


\vspace{-1mm}
\section{Introduction}
\label{sec:intr}



In the past few years, a lot of attention is paid to ``training-on-edge'', which is highly promoted by intelligent edge applications.
	As one of the most well-recognized edge training techniques, Federated Learning (FL) expects to unit multiple resource-constrained edge devices to collaboratively train identical neural network models with their local dataset~\cite{zhao2018federated}.
	By aggregating the parameter updates from each device, a global model can be collaboratively trained efficiently and securely.

\begin{figure}[t]
	\centering
	\captionsetup{justification=centering}
	\vspace{0mm}
	\includegraphics[width=3.3in]{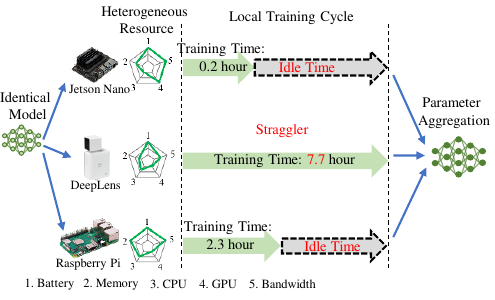}
	\vspace{-2mm}
	\caption{The Straggler Issue in Original FL}
	\vspace{-6mm}
	\label{straggler}
\end{figure}

When applying FL into practical edge training, a serious problem emerges:
Although edge devices train the same model structure and identical workload, they may have distinct computation resources (\textit{e.g.}, memory size, CPU/GPU bandwidth, \textit{etc}).
	Therefore, like the ``shortest board in barrel'', the devices with extremely weak computation capacities will take much longer time for local training.
	These devices are referred to as \textit{stragglers} in FL\@.
	One example is demonstrated by Fig.~\ref{straggler}: with synchronous FL aggregation setting, stronger collaboration nodes (\textit{Jetson Nano} and \textit{Rasberry Pi}) have to keep an idle status to wait for the straggler (\textit{DeepLens}) in every cycle and prolong the training cycle from 2.3 to 7.7 hours.

Intuitively, changing the synchronous FL aggregation into an asynchronous manner can directly kick out the stragglers from most aggregation cycles and accelerate the training cycles~\cite{wu2019safa,chen2019asynchronous,chen2019communication,mohammad2019adaptive,xie2019asynchronous}.
	However, in FL settings where most local dataset are non-identical distributed (Non-IID), the stragglers may learn unique and critical information.
	And dramatically asynchronous collaboration may defect both local and global convergence performance~\cite{hakimi2019taming,zhang2015staleness}.


To solve the FL straggler issue, we propose \textit{Helios}, a heterogeneity-aware FL framework.
	\textit{Helios} identifies the stragglers regarding the collaborative training pace and specify the expected neural network model training volumes.
	For straggling devices, a “soft-training” method is proposed to dynamically compress the original identical training model into the expected volume through a rotating neuron training approach.
	With extensive algorithm analysis and optimization schemes, the stragglers can be accelerated while retaining the convergence for local training and federated collaboration.

Our paper has the following three main contributions:
\begin{itemize}
	\item The straggler identification methodology we proposed incorporates different approaches (\textit{i.e.}, time-based approximation and resource-based profiling) for various FL deployment contexts, offering outstanding feasibility for FL straggler research works;
	\item A collaborative training method --- ``soft-training'' is proposed to keep different part of model parameters on the stragglers alternately join the training cycles, which not only maintains synchronized FL with accelerated stragglers to preserve learning information integrity from every node, but also preserve straggler model integrity without unrecoverable compression;
	\item Dedicated optimization schemes are proposed based on extensive algorithm analysis and proof to enhance the overall performance.
\end{itemize}

Experiments show that \textit{Helios} can provide up to 2.5$\times$ training speed-up and maximum 4.64\% convergence accuracy improvement in various FL settings with stragglers.

\vspace{-1mm}
\section{Preliminary}
\label{sec:prel}
\vspace{-1mm}

\subsection{\textbf{Federated Learning on Heterogeneous Edge Devices}}
\vspace{-1mm}

In practical FL implementation, multiple edge devices collaboratively train the identical models on their local training data with two kinds of heterogeneities:

\textit{\textbf{Hardware Resource Heterogeneity:}}
Since various edge devices participate in FL, they have heterogeneous hardware resources in terms of memory size, CPU bandwidth, and \textit{etc}.
Given a certain edge device in FL, the training consumption of its CNN model can be generally evaluated as the training cycle time cost.
However, the mismatches between the model's computation consumption and device's resource capacities will increase the model training time. 


\textit{\textbf{Information Heterogeneity:}}
	In FL process, every edge device will constantly learn the information from local training data and contribute to the final global model~\cite{nasr2018comprehensive,yang2018applied}.
Although lagging the FL efficiency, stragglers still have important learned information. 
This phenomenon becomes more prominent in the Non-IID setting~\cite{zhao2018federated}. 
The critical information learned from stragglers are distinct and cannot be ignored from the perspective of the entire collaboration process.

Based on the above two heterogeneity analysis, it is essential to eliminate stragglers delay while preserving their learned information in FL process.

\begin{figure}[t]
	\centering
	\captionsetup{justification=centering}
	\vspace{-8mm}
	\includegraphics[width=3.5in]{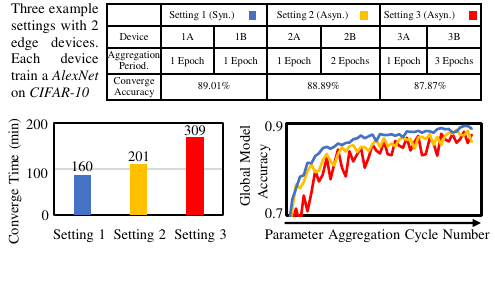}
	\vspace{-15mm}
	\caption{Asynchronous FL Performance Evaluation}
	\vspace{-8mm}
	\label{syn_vs_asyn}
\end{figure}

\vspace{-2mm}
\subsection{\textbf{Federated Learning Straggler Issue}}
\vspace{-1mm}

Some works are proposed to solve the straggler issue:

\textit{\textbf{Asynchronous Methods}}: these works only aggregate parameters of stragglers at certain cycles~\cite{nishio2019client,caldas2018expanding,wang2019adaptive,chen2019asynchronous}.
	Nishio \textit{et al.} proposed an optimized FL protocol (\textit{i.e.}, FedeCS) to kick out straggled devices from the learning collaboration~\cite{nishio2019client}.
	Wang \textit{et al.} introduced a dedicated collaboration method to reduce the training loss introduced by the asynchronous stragglers~\cite{wang2019adaptive}. 
Although showing acceleration performance, they cannot fundamentally eliminate stragglers and even cause information degradation and stale parameter updating ~\cite{hakimi2019taming,zhang2015staleness}.
Fig.~\ref{syn_vs_asyn} shows an analysis for two collaborative devices under three settings.
	We can easily find that, the original synchronous FL achieves the best convergence accuracy.
	However, when the asynchronous straggler parameter aggregation cycle increases from 2 epochs (setting 2) to 3 epochs (setting 3), both the converge accuracy and speed will decrease.

\begin{figure}[t]
	\centering
	\captionsetup{justification=centering}
	\vspace{-10mm}
	\includegraphics[width=3.3in]{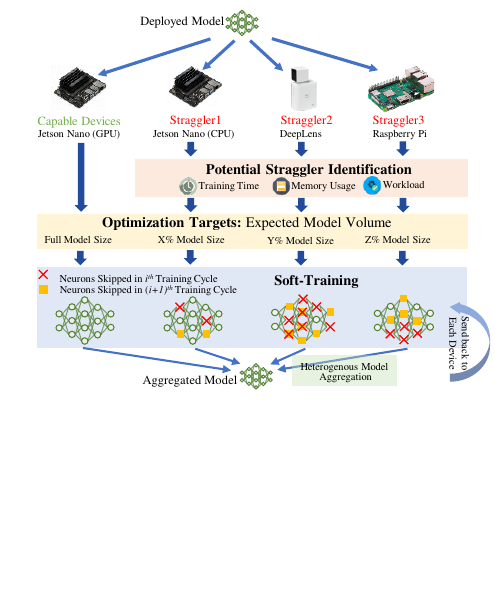}
	\vspace{-37mm}
	\caption{The \textit{Helios} Framework Overview}
	\vspace{-7mm}
	\label{overview}
\end{figure}

\textit{\textbf{Synchronous Methods}}: these works leveraged local training optimization methods to accelerate stragglers' training cycles so as to achieve synchronous aggregation~\cite{jiang2019model,jeong2018communication}.
	Jiang \textit{et al.} used model pruning to compress the original models to smaller ones which can satisfy the given constraints~\cite{jiang2019model}.
	Jeong \textit{et al.} reduced the model size in edge nodes based on knowledge distillation~\cite{jeong2018communication}.
However, due to the permanent model structure losing, the information capacity of stragglers will decrease and may face converge degradation.
	As shown in~\cite{yuan2019distributed}, when introducing models with diverged structures, the global model accuracy would drop as much as 10\%.


\vspace{-2mm}
\section{Framework Overview} 
\label{sec:recon}
\vspace{-1mm}

The overall concept of \textit{Helios} is shown in Fig.~\ref{overview}.
In order to accelerate stragglers, the proposed framework is aiming to optimize the model training on stragglers.

\textit{\textbf{Initialized Stragglers Identification:}}
Since the straggler issue is caused by devices' heterogeneous resources, we proposed two straggler identification approaches for various FL deployment contexts to improve framework flexibility, which will be discussed in Section IV. B.




\textit{\textbf{Optimization Target Determination:}}
In Section IV. C, the initialized stragglers' optimization targets (\textit{i.e.} the expected model volumes) are further determined, which can reduce models' training consumption and achieve acceleration. 


\textit{\textbf{Soft-training:}}
Guided by the expected model structure volumes, ``soft-training'' method is introduced. In each training cycle, the proposed method can let different part of model parameters (\textit{e.g.} neurons) alternately join the training, which maintains a complete model parameter updating for collaboration.
As Fig.~\ref{overview} shows: in the $i^{th}$ training cycle, neurons indicated by red crosses are skipped to achieve training acceleration. However, in the  $(i+1)^{th}$ training cycle, a different set of neurons (yellow squares) are skipped.
Such a training method can guarantee each model parameter on stragglers has opportunities to make a contribution to the collaboration.



\textit{\textbf{Optimizations for Soft-training:}}
We further propose several dedicate optimization schemes in Section VI, which can enhance the overall convergence performance and collaboration scalability based on the extensive algorithm analysis and proof.


\vspace{0mm}
\section{Potential Straggler Identification in Heterogeneous Collaboration}
\label{sec:prof}
\vspace{-1mm}


In this section, we will discuss potential straggler identification and the corresponding optimization target determination in the heterogeneous collaboration setting.

\vspace{-1mm}
\subsection{\textbf{Heterogeneous Federated Learning Setup}}
\vspace{-1mm}

In the heterogeneous FL, multiple edge devices with heterogeneous computation resources collaboratively train the identical models with their local datasets. After each local training, they upload the local model parameters to the global device for global model updating.
Once obtaining a new global model, the global device sends the model parameters back to each local device. During this process, the devices with extremely weak computation capacities will take a much longer local training time, becoming stragglers in FL.

\vspace{-1mm}
\subsection{\textbf{Straggler Identification}}
\vspace{-1mm}

Since the straggler has a much longer training time cost than the normal devices (35$\times$ in Fig.~\ref{straggler}), identifying it only after FL process will introduce a high time overhead.
Therefore, the potential stragglers need to be initially detected by comparing the computation resources of each device. Two approaches are discussed in this paper: time-based approximation and resource-based profiling. 

\textit{\textbf{Time-based Approximation:}}
If partial/all devices' hardware configurations cannot be obtained (\textit{\textbf{Black Box}}), we can leverage the proposed time-based approximation to fast identify the stragglers.
In this case, our approach will assign each device with a lightweight test bench (only train few iterations) and quickly record their training time cost. 
Then the approach ranks all devices based on their training time cost values and obtain an approximate index $T = \{T_1, T_2, ..., T_N\}$, where $T_1$ represents the longest time cost.
Based on FL requirement, the top-k devices in the index are identified as potential stragglers.


\textit{\textbf{Resource-based Profiling:}}
If all devices' hardware configurations can be obtained (\textit{\textbf{White Box}}), we can leverage the proposed resource-base profiling to accurately identify the stragglers. 
First, the scheme will investigate the computation resource of each device in terms of computation bandwidth, memory size, communication bandwidth, \textit{etc.}
Next, we can leverage resource models in~\cite{xu2020directx} to fully profile each device's model training consumption.
For example, straggler's training time cost $Te$ can be exactly formulated by the training computation workload $W$, memory usage $M$ as: \small $Te= W/C_{cpu} + M/V_{mc} + M/B_n$\normalsize, where $C_{cpu}$, $V_{mc}$, and $B_n$ represent device computation bandwidth, data transmitting speed, and communication bandwidth.
Finally, we can identify the stragglers based on each device's computation ability.


\vspace{-1mm}
\subsection{\textbf{Optimization Target Determination}}
\vspace{-1mm}


After finding the potential stragglers, it is important to identify the optimization target, namely, the expected model volumes of stragglers in each training cycle to achieve acceleration. 
The expected model volume can be selected with a pre-define volume or adapted to a specific value according to the resource constraints.

For the first one, we can define multiple model volume levels in advanced and assign each straggler with a model volume level according to the training time cost index $T$. 
The model volume will be dynamically adjusted to an optimal point during the first several training cycles in FL process. 

If the model volumes of stragglers need to be accurately determined before the training, we can leverage the profiling models we built in the \textit{Resource-based profiling}.
Specifically, we select each layer with $P_i n_i$ neurons simultaneously until the model consumption approaches to the resource constraints, where $n_i$ is the total neuron number in the $i^{\textit{th}}$ layer and $P_i$ is a ratio between $(0,1)$. 
Hence, the optimized model volumes can ensure training acceleration during the FL process.
\vspace{-1mm}
\section{Soft-training on Straggling Edge Devices } 
\label{sec:recon}

With identified potential straggler and corresponding model volumes from initialization, we propose ``soft-training'' scheme to adapt the optimized model to specific resource constraints while guaranteeing converge effectiveness.

\vspace{-1mm}
\subsection{\textbf{Soft-training Process under Optimization Targets}}
\vspace{-1mm}

\begin{figure}[t]
	\centering
	\captionsetup{justification=centering}
	\vspace{0mm}
	\includegraphics[width=3.3 in]{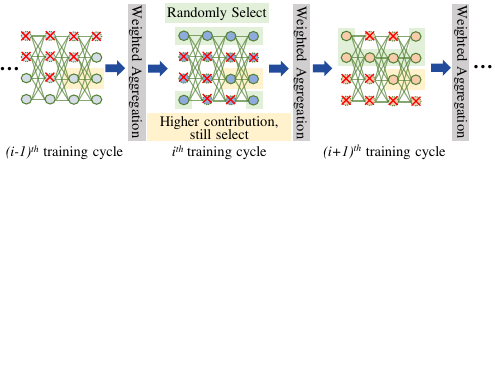}
	\vspace{-37mm}
	\caption{Soft-Training Scheme Flow}
	\vspace{-6mm}
	\label{soft_training}
\end{figure}

Fig.~\ref{soft_training} shows the entire ``soft-training'' flow and we take neurons as our minimum model parameter structure.
The entire ``soft-training'' can be divided into three steps: 

\textit{\textbf{Step 1. Straggler Model Shrinking:}}
In order to meet the optimization targets (the expected model volume), we only select a set of neurons from stragglers' models to join the training cycle.
As aforementioned, methods in Section IV only approximately identify stragglers and their expected mode volume.
Therefore, \textit{Helios} needs first few training cycles to finalize the stragglers and model volumes by dynamically adjusting the model volumes to accommodate normal devices \textit{w.r.t.} training time cost.

During the partial training, two types of neurons are selected to achieve an optimized convergence performance: the neurons with the higher contributions to the collaboration convergence (primary converge guarantee) and some other random neurons (further converge optimization). 
Then we define the contribution metric by leverage the assumption in~\cite{alistarh2018convergence} that the neurons with larger weight parameter changing values will provide larger impacts to the global model.
We assume the weight parameters of the $j^{\textit{th}}$ neuron in the $i^{\textit{th}}$ layer after training epoch $S_k$ as $\theta_{s,r,n}^{ij}(S_k)$, and the neuron's collaboration contribution $U^{ij}(S_k)$ will be calculated by the summation of useful changes in the current training cycle as: 

\vspace{-1mm}
\scriptsize
\begin{equation}
	\medmuskip=-2mu
	\thinmuskip=-2mu
	\thickmuskip=0mu
	\hspace{-1mm}
	U^{ij}(S_k)= \theta_{s,r,n}^{ij}(S_k) - \theta_{s,r,n}^{ij}(S_{k-1}),
	\label{eq:neuroncontribution}
	\vspace{-1mm}
\end{equation}
\normalsize
where a larger $U^{ij}$ represents the target neuron demonstrates a higher collaboration contribution.
We will elaborately show how to select these two types of neurons in each training cycle in Step 2 and Step 3.


\textit{\textbf{Step 2. Neuron Rotation for Model Integrity:}}
With partial training in Step 1, model training consumption on stragglers can be adapted to the specific device resource constraints. 
Some previous methods leverage fixed model pruning to achieve partial training~\cite{jiang2019model,jeong2018communication}.
However, without an optimized training algorithm to keep a rather complete model structure, the pruned model parameters will permanently lose their contribution to the collaboration, thereby inevitably hurt convergence accuracy and speed.


Therefore, in our ``soft-training'', we solve the conflict between the optimization targets and model integrity by letting different part of model parameters (\textit{i.e.} neurons) alternately join training cycles.
As Fig.~\ref{soft_training} illustrates, in the $i^{th}$ training cycle, we first choose $P_s$ percentage neurons\footnote{We empirically evaluate overhead of such sorting operation and find it can be ignored compared to the training cost (\textit{e.g.} 18ms vs 12mins). } which have the highest changing values $U^{ij}$ and randomly select other $(1-P_s)P_i n_i$ neurons from the rest $(n_i-P_s P_i n_i)$ ones which have relative lower contributions. Value of $P_s$ will be discussed in Section VI. A
Therefore, the total parameter changing on a straggler is:

\vspace{-2mm}
\scriptsize
\begin{equation}
	\medmuskip=-2mu
	\thinmuskip=-2mu
	\thickmuskip=0mu
	\hspace{-1mm}
	\Theta(S_k)-\Theta(S_{k-1}) = TopK(U^{ij}) \ \cup \ Rand(U^{ij}), \ where\ K = P_s P_i n_i,
	\label{eq:neuroncontribution2}
	\vspace{-1mm}
\end{equation}
\normalsize
where $\Theta(S_k)$ represent the model parameter of the straggler and $Rand$ operation is randomly selecting $(1-P_s)P_i n_i$ neurons' $U^{ij}$. 
Next, in the $(i+1)^{th}$ training cycle, since the neurons with the highest contribution don't change, we still select them in the training to provide the primary converge guarantee. 
On the contrary, a different set of neurons with lower contributions will be selected. 
By alternately joining each training cycle, every neuron on a straggler has opportunities to keep model integrity to the global model convergence.

\textit{\textbf{Step 3. Aggregation and Local Model Updating:}}
After finishing a local training cycle, each local edge device will upload the gradients of the selected neurons to the global device during the aggregation cycle.
    Meanwhile, the global device will average the uploaded model parameters to update the global model, which will be further discussed in Section VI. B. 
Once the new global model parameters are obtained, they will be sent to each local device for local model updating and the local edge devices can start the next training cycle.

With the above three steps, models on the stragglers adapted to the optimization targets while still maintain a rather complete model structure and balanced contribution, thereby guarantees the global model convergence performance.

\vspace{-1mm}
\subsection{\textbf{Convergence Proof and Conditions Analysis}}
\vspace{-1mm}

Consider federated learning with $N$ nodes, the training dataset and loss function for each device is $\left\{x_{n}\right\}$ and $\left\{f_{n}\right\}$, respectively. Let $\Theta \in \mathbb{R}^m$  represents the model parameter and $m$ indicates the total neuron number. 

The FL optimization goal can be formulated as:

	\vspace{-2mm}
	\scriptsize
	\begin{equation}
	\medmuskip=-2mu
	\thinmuskip=-2mu
	\thickmuskip=0mu
	\hspace{-1mm}
	\min _{\Theta} \quad f(\Theta):=\frac{1}{N} \sum_{n=1}^{N} f_{n}(\Theta), \quad \Theta_{t+1}=\Theta_{t}-\eta_{t} g_{t}\left(\Theta_{t}\right),
	\vspace{-2mm}
	\label{eq8}
	\end{equation}
	\normalsize
where $t$ is the current training epoch number (equals to $S_{k}$ in Eq.~\ref{eq:neuroncontribution2}), and $g_{t}(\cdot)$ represents the gradients.
%
%

\vspace{1mm}
\noindent \textbf{Proposition-1}: \textit{Convergence Bounded by Gradient Variance}. 
Based on previous work~\cite{wangni2018gradient}, the global model convergence loss is proved to be bounded by the variance of gradient $g_{t}\left(\Theta_{t}\right)$ across different nodes:

    \vspace{-1mm}
    \scriptsize
	\begin{equation}
	\mathbb{E}\left[f\left(\Theta_{t+1}\right)\right] \leq f\left(\Theta_{t}\right)-\eta_{t}\left\|\nabla f\left(\Theta_{t}\right)\right\|^{2}+\frac{L}{2} \eta_{t}^{2} {\mathbb{E}\left\|g_{t}\left(\Theta_{t}\right)\right\|^{2}}.
	\vspace{-1mm}
	\end{equation}
	\normalsize

Therefore, in our soft-training algorithm, we regulate the third term, which is the gradient variance to achieve the desired model convergence. 
    Such theoretical convergence boundary can be also applied to Non-IID situation, where the gradient variance ${\mathbb{E}\left\|g_{t}\left(\Theta_{t}\right)\right\|^{2}}$ can still find a upper boundary.

\vspace{1mm}
\noindent \textbf{Proposition-2:} \textit{Bounded Gradient Variance in Soft-Training}.
The gradient variance in soft-training is bounded by $(1+\rho) v$ \textit{if we can maintain at least $v$ neurons with the highest gradient contribution, thus ensuring the convergence of soft-training.}

\vspace{1mm}
\noindent \textbf{Proof:}
In soft-training, we select partial neurons to join each training cycle for acceleration. 
Denote the neurons' gradient in soft-training as $ST\left(g\left(\Theta_{t}\right)\right)$ and simplify $g_{t}\left(\Theta_{t}\right)$ as $g$. 
Each neuron's $g$ has a probability $p_{i}$ to be selected in each cycle. 
Therefore, the gradient variance could be reformulated as: 

    \vspace{-1mm}
    \scriptsize
	\begin{equation}
	\scriptsize
	ST(g)=\left[D_{1} \frac{g_{1}}{p_{1}}, D_{2} \frac{g_{2}}{p_{2}}, \cdots, D_{d} \frac{g_{d}}{p_{d}}\right],
	\vspace{-1mm}
	\label{eq:st_grad}
	\end{equation}
	\normalsize
where $D_{i} \in\{0,1\}$ is the mask indicating the selection results of each neuron, and each gradient term $g_i$ divided by probability $p_i$ denotes the unbiased estimation of $g_i$. \textit{Moreover, $p_i$ cannot be 0 therefore each neuron shouldn't be inactivated for a long-term.}
Thus, the variance of $ST(g)$ could be then reformulated as:

    \vspace{-1mm}
    \scriptsize
	\begin{equation}
	\scriptsize
	\mathbb{E} \sum_{i=1}^{m}\left[ST(g)_{i}^{2}\right]=\sum_{i=1}^{m}\left[\frac{g_{i}^{2}}{p_{i}^{2}} \times p_{i}+0 \times\left(1-p_{i}\right)\right]=\sum_{i=1}^{m} \frac{g_{i}^{2}}{p_{i}},
	\vspace{-1mm}
	\end{equation}
	\normalsize
which shows the variance of gradient is related to the probability $p_{i}$. Therefore,
the trade-off between $p_{i}$ and gradient variance could be formulated as an optimization problem:

    \vspace{-1mm}
    \scriptsize
	\begin{equation}
	\scriptsize
	\min _{p} \sum_{i=1}^{m} p_{i} \quad \text { s.t. } \quad \sum_{i=1}^{m} \frac{g_{i}^{2}}{p_{i}} \leq(1+\epsilon) \sum_{i=1}^{m} g_{i}^{2}, 
	\vspace{-1mm}
	\end{equation}
	\normalsize
where $\epsilon$ controls the variance of gradient.


\vspace{1mm}
\noindent \textbf{Condition of Convergence}. The convergence of soft-training could then be ensured with the following neuron selection criteria.
During each soft-training cycle, we keep the $P_{s} n_{c}$ neurons with highest convergence contributions joining the next training cycle. 
Let $C$ represents the set of such high contribution neurons and their total number is $v$, i.e., $p_{i}=1$ for $i \leq v$. In order to determine the value of $v$, we have:


    \vspace{-1mm}
    \scriptsize
	\begin{equation}
	\scriptsize
	\sum_{i=1}^{m} \frac{g_{i}^{2}}{p_{i}}-(1+\epsilon) \sum_{i=1}^{m} g_{i}^{2} 
	=\sum_{i=1}^{v} g_{(i)}^{2}+\sum_{i=v+1}^{m} \frac{\left|g_{(i)}\right|}{\lambda}-(1+\epsilon) \sum_{i=1}^{m} g_{i}^{2}=0.
	\vspace{-1mm}
	\end{equation}
	\normalsize
Thus according to ~\cite{wangni2018gradient}, we can guarantee the convergence with $v$ and $\rho = \epsilon$ and
$\mathbb{E}\left[\|ST(g)\|_{0}\right]$ is bounded by:

    \vspace{-1mm}
    \scriptsize
	\begin{equation}
	\begin{aligned} 
	\scriptsize
	&\mathbb{E}\left[\|ST(g)\|_{0}\right] =\sum_{i=1}^{m} p_{i}=\sum_{i \in C_{v}} p_{i}+\sum_{i \notin C_{v}} p_{i} \\ 
	& \leq v+\frac{\rho^{2} v\left\|g_{C_{v}}\right\|_{2}^{2}}{\rho\left\|g_{C_{v}}\right\|_{2}^{2}+(1+\rho)\left\|g_{C_{v}}\right\|_{2}^{2}} \leq(1+\rho) v. 
	\vspace{-1mm}
	\end{aligned}
	\label{eq:9}
	\end{equation}
	\normalsize
Applying the aforementioned dynamic neuron selection, the gradients in the soft-training will be bounded by Eq.~\ref{eq:9} and thus soft-training convergence could be achieved.

\vspace{-1mm}
\section{Optimizations for Soft-training} 
\label{sec:recon}

In this section, we propose several schemes to enhance overall performance based on the algorithm analysis and improve collaboration scalability. 

\vspace{-1mm}
\subsection{\textbf{Neuron Rotation Regulation}}
\vspace{-1mm}

According to the proposition-2 and the corresponding condition, the gradient variance of soft-training is bounded by $P_s$, which is the ratio of neurons with the highest contribution $U^{ij}$. If $P_s = 1$, $\rho$ will equal 0, which means the local training with full neurons on a normal device. According to the previous works~\cite{alistarh2018convergence, lin2017deep}, $P_s$ usually can be selected from 0.05 to 0.1 for stragglers during FL process. 

Another condition requirement is: neurons should not be inactivated for the long-term, which may introduce 0 for $p_i$ and stale parameter issue.
In order to avoid this, we pull the long-term skipped neurons back to training cycles timely.
Specifically, during aggregation, each straggler will send its partial updated gradients index to the global device. The global device will record all currently skipped neurons and update their skipped cycles value $C_s$. In every aggregation cycle, once a certain neuron's $C_s$ exceeds a pre-define threshold (we set as $1+\frac{m}{\sum p_i n_i}$), the global device will inform the corresponding straggler and the targeted neuron will rejoin the training.

\begin{table}[t!]
\vspace{-8mm}
	\small
	\centering
	\vspace{-1.5mm}
	\caption{4 Stragglers with Heterogeneous Resource}
		\vspace{-1.5mm}
	\setlength{\tabcolsep}{1.5mm}{
\begin{tabular}{|c|c|c|c|c|}
\hline
Constraints      & \begin{tabular}[c]{@{}c@{}}Nano\\ (CPU)\end{tabular} & Raspberry & \begin{tabular}[c]{@{}c@{}}Deeplen\\ (GPU)\end{tabular} & \begin{tabular}[c]{@{}c@{}}DeepLen\\ (CPU)\end{tabular} \\ \hline
Comp. W (GFLOPS) & 7                                                    & 6         & 5.5                                                     & 4.5                                                     \\ \hline
Mem. U (MB)      & 252                                                  & 150       & 100                                                     & 110                                                     \\ \hline
Tim. C (Mins)    & 20.6                                                 & 23.8      & 27.2                                                    & 34                                                      \\ \hline
\end{tabular}}
	\begin{tablenotes}
		\scriptsize
		\item[1] \hspace{-2mm} Comp. W: Computation Workload; Mem. U: Memory Usage; Tim. C: Time cost;
	\end{tablenotes}
		\vspace{-6mm}
	\label{tab:exp}
\end{table}
\normalsize


\vspace{-1mm}
\subsection{\textbf{Model Aggregation Optimization with Heterogeneity}}
\vspace{-1mm}

From the perspective of the overall collaboration process, the proposed soft-training can maintain a balanced model updating of the stragglers.
However, dive into each parameter aggregation cycle, multiple stragglers will introduce various partial models with diverged structures.
Such a limited model size will bring additional errors to the global model. 
Hence, we introduce heterogeneous model aggregation scheme: according to the updating model size of device, assign each device with different weights to adjust their contribution:

\vspace{-1mm}
\scriptsize
\begin{equation}
	\min _{\Theta} \quad f(\Theta):=\frac{1}{N} \sum_{n=1}^{N} \alpha_{n}f_{n}(\Theta)
	\vspace{-1mm}
\end{equation}
\normalsize
where $\alpha_{n}$ is the adjusting ratio, $\sum\alpha_{n}=1$ and $\alpha={r_{n}}/{\sum r_{n}}$. $r_{n}$ is the ratio of neurons selected on each device. A larger $\alpha_{n}$ means a more complete model structure will have more contributions to the collaboration performance.

\vspace{-2mm}
\subsection{\textbf{Collaboration Scalability Optimization}}
\vspace{-1mm}

During the practical FL process, more devices will dynamically join the collaboration.
We leverage the following steps to improve the collaboration scalability of our method: 
Once \textit{Helios} detecting a new device is adding into FL, it will first fetch device's hardware computation resources and compare them with the existed stragglers. During this step, both approximation and profiling approaches can be used to identify the new adding device based on whether the FL has enough profiling budgets. 
If the new device is identified as a straggler, \textit{Helios} will assign it with a pre-defined model volume or adapt its model size based on the resource constraints.

\vspace{-1mm}
\section{Experiment}
\label{sec:expe}
\vspace{-1mm}

\subsection {\textbf{Experiment Setup}}
\vspace{-1mm}

\begin{figure}[!t]
	\centering
	\captionsetup{justification=centering}
	\vspace{-11mm}
	\includegraphics[width=3.2in]{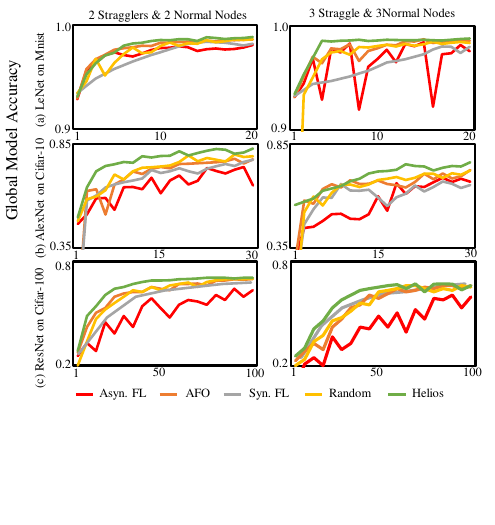}
	\vspace{-20mm}
	\caption{Soft-training Effectiveness Evaluation}
	\vspace{-7mm}
	\label{scalability}
\end{figure}

\textbf{\textit{Testing Platform Setting:}}
In our experiment, we leverage resource-based profiling to identify the stragglers. We first profile the several edge devices' resource configurations. 
Then, by adjusting the configuration of CPU/GPU bandwidth and memory availability, we can simulate these edge devices' training performance on multiple Nvidia Jetson Nano development boards and let them as stragglers with different resource capabilities. 
	The details of these straggler settings when running \textit{AlexNet} on \textit{CIFAR-10} are shown in Table.~\ref{tab:exp}.

\textbf{\textit{CNN Models and Dataset:}}
In the experiment, three CNN models are used as our testing targets, namely, \textit{LeNet}, \textit{AlexNet}, and \textit{ResNet-18}.
	The above three CNN models are trained on \textit{MNIST}, \textit{CIFAR-10}, and \textit{CIFAR-100}, respectively.

\textbf{\textit{Baseline Methods:}}
In order to exhibit the superiority of \textit{Helios}, we re-implement five other FL schemes for comparison:
  (1) \textit{Synchronized FL (Syn. FL)}: All devices (including stragglers) update their parameters synchronously.
  (2) \textit{Asynchronous FL (Asyn. FL)} Normal edge devices update their parameters immediately after local training without waiting for stragglers.
  (3) \textit{Random}~\cite{caldas2018expanding}: In each training, the stragglers randomly select partial model with the expected model structure volume.
  (4) \textit{AFO}~\cite{xie2019asynchronous}: A optimized asynchronous method aiming to reduce staleness issue of stragglers.

\vspace{-1mm}
\subsection {\textbf{General Helios Performance Evaluation }}
\vspace{-1mm}

We evaluate \textit{Helios} in terms of the converge accuracy and speed by comparing the above baselines in this part.
	As shown in Table.~\ref{tab:exp}, there are two straggler settings involved: (1) Four devices join in the FL with two capable devices and two straggled devices as Strag. 1 and Strag. 2.
	(2) Six devices join in the FL with three capable devices and three straggled devices as Strag. 1, Strag. 2 and Strag. 3.

\textbf{\textit{Accuracy Evaluation.}}: Fig.~\ref{scalability} illustrates the experimental results on converge accuracy. The X-axis represents the aggregation cycles of capable edge devices.
We can find that \textit{Asyn.~FL} always achieves the lowest accuracy due to the information degradation. 
As for \textit{Syn.~FL}, since its training cycle is determined by the stragglers, it shows a much lower aggregation speed than other methods. 
\textit{Helios} always shows the best accuracy compared to all the baseline methods. 
Specifically, for \textit{LeNet}, \textit{AlexNet}, and \textit{ResNet18}, \textit{Helios} achieves at most 0.24$\%$ to 4.64$\%$ accuracy improvement on two kinds of straggler settings.

\textbf{\textit{Convergence Speed Evaluation}}: 
It is clear that \textit{Asyn.~FL} has the worst converge speed, and even fail to converge due to the staleness parameters and information loss. 
Although \textit{AFO} optimizes the asynchronous updates, its convergence is still affected by the staleness parameters of stragglers.
On the contrary, \textit{Helios} shows the fastest converge speed, especially for the FL with more stragglers. For the experiments of 6 nodes on \textit{MNIST}, \textit{CIFAR-10}, and \textit{CIFAR-100}, \textit{Helios} approaches convergence after 4, 12, and 40 aggregation cycles respectively.
	On the contrary, other methods need at least 10, 18, and 50 aggregation cycles.
Overall, \textit{Helios} can achieve at most 2.5$\times$ speedup than the state-of-the-art methods.

\begin{figure}[!t]
	\centering
	\captionsetup{justification=centering}
	\vspace{-7mm}
	\includegraphics[width=3.4in]{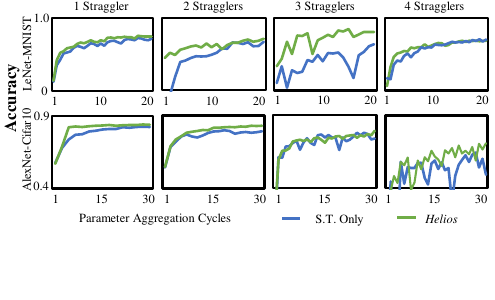}
	\vspace{-13mm}
	\caption{Model Aggregation Optimization Evaluation}
	\vspace{-5mm}
	\label{exp3}
\end{figure}

\vspace{-1mm}
\subsection {\textbf{Model Aggregation Optimization Evaluation}}
\vspace{-1mm}

In this part, the effectiveness of model aggregation optimization scheme is also evaluated under the same setting (stragglers are increasing from 1 to 4). We use soft-training without aggregation optimization as the baseline, which is represented as \textit{S.T. Only} .
Fig.~\ref{exp3} illustrates the comparison results between \textit{Helios} and \textit{S.T. Only}. 
We can see that \textit{Helios} achieves at most 17.37\% accuracy improvement and reduce the accuracy variance caused by partial model aggregation while accuracy curve of \textit{S.T. Only} still has obvious fluctuations. 

\vspace{-1mm}
\subsection {\textbf{Non-IID Setting Evaluation}}
\vspace{-1mm}

\begin{figure}[!t]
	\centering
	\captionsetup{justification=centering}
	\vspace{-7mm}
	\includegraphics[width=3.3in]{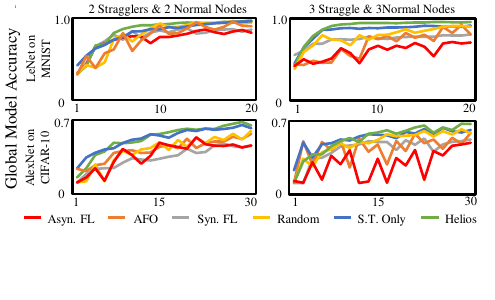}
	\vspace{-11.5mm}
	\caption{\textit{Helios} Evaluation with Non-IID Data}
	\vspace{-5mm}
	\label{uneqnon}
\end{figure}

In this part, we evaluate the effectiveness of \textit{Helios} with a Non-IID setting (though our proposed technique is more oriented by computational heterogeneity rather than data heterogeneity).
Specifically, Non-IID capability is evaluated with 4 and 6 edge devices with 2 and 3 stragglers, respectively.
We use the same Non-IID data generation method in~\cite{zhao2018federated}.
Fig.~\ref{uneqnon} shows the evaluation results. 
We can find that Non-IID data brings performance degradation for all methods. 
However, compared with other methods, \textit{Helios} can always obtain a better converge accuracy and speed.
	\vspace{-1mm}
\section{Conclusion}
\label{sec:conc}
	\vspace{-1mm}

In this work, we proposed \textit{Helios} --- a heterogeneous-aware FL framework with dynamically balanced collaboration.
Leveraging on-device model training consumption profiling and the innovative ``soft-training'' training scheme, \textit{Helios} can introduce effective local CNN model optimization into FL to eliminate stragglers, while maintaining expected collaborative convergence across all edge devices.
Experiments demonstrated that the proposed \textit{Helios} achieves superior training accuracy, speed, as well as Non-IID setting resistance.
\textit{Helios} significantly enhanced the applicability and performance of federated learning for training-on-edge.

\hspace{-1mm}
\textbf{Acknowledgment:}
\vspace{-1mm}
This work was supported in part by NSF CNS-2003211.
\vspace{-2mm}

\bibliographystyle{IEEEtran}
\scriptsize
\vspace{-1mm}
\bibliography{DAC_ELFISH}


\end{document}